\documentclass[prl,twocolumn,superscriptaddress,floatfix,showpacs,showkeys]{revtex4}

\usepackage{graphics}% Include figure files
\usepackage{times}
\newcommand{\ie}{{\it i.e.}}
\newcommand{\eg}{{\it e.g.}}
\newcommand{\cf}{{\it c.f.}}
\newcommand{\vs}{{\it vs.}}

\newcommand{\YBCO}{YBa$_2$Cu$_3$O$_7$}

\newcommand{\Optional}[1]{%
\newlength{\OLen}\setlength{\OLen}{\columnwidth}%
\addtolength{\OLen}{-2\fboxsep}%
\par\noindent\fcolorbox{black}{yellow}{%
\parbox{\OLen}{{\small\sffamily To skip}}}%
\par\noindent\fbox{\parbox{\OLen}{{\small\sffamily #1}}}}

\newcommand{\units}[1]{\,{\rm #1}}

\renewcommand{\Optional}[1]{}

\renewcommand{\section}[2][]{}
\renewcommand{\subsection}[2][]{}

\begin{document}

\title{
  Dynamics of semifluxons in Nb long Josephson 0-$\pi$ junctions
}

\author{E.~Goldobin}
\email{gold@uni-tuebingen.de}
%\homepage{http://www.geocities.com/e_goldobin}
\affiliation{
  Physikalisches Institut, Experimentalphysik II,
  Universit\"at T\"ubingen,
  Auf der Morgenstelle 14,
  D-72076 T\"ubingen, Germany
}
\affiliation{
  Institut f\"ur Mikro- und Nanoelektronische Systeme,
  Universit\"at Karlsruhe (TH),
  Hertzstrasse 16,
  D-76187 Karlsruhe, Germany
}

\author{A.~Sterck}
\author{T.~Gaber}
\author{D.~Koelle}
\author{R.~Kleiner}
\affiliation{
  Physikalisches Institut, Experimentalphysik II,
  Universit\"at T\"ubingen,
  Auf der Morgenstelle 14,
  D-72076 T\"ubingen, Germany
}



%Acronims:
% LJJ:66, SF, PSF, NSF, AFM:20, ZFS:9
%\definecolor{gray}{gray}{0.75}
%\date{9.9.2002 [\colorbox{gray}{cond-mat/0209214}]}
\date{\today}

\begin{abstract}
  We propose, implement and test experimentally long Josephson 0-$\pi$ junctions fabricated using conventional Nb-AlO$_x$-Nb technology. We show that using a pair of current injectors, one can create an arbitrary discontinuity of the Josephson phase and in particular a $\pi$-discontinuity, just like in $d$-wave/$s$-wave or in $d$-wave/$d$-wave junctions, and study fractional Josephson vortices which spontaneously appear. Moreover, using such junctions, we can investigate the \emph{dynamics} of the fractional vortices --- a domain which is not yet available for natural 0-$\pi$-junctions due to their inherently high damping.  We observe half-integer zero-field steps which appear on the current-voltage characteristics due to hopping of semifluxons.
\end{abstract}

%{\small To be submitted to Phys.\ Rev.\ Lett. (4 pages).}

\maketitle

\section{Introduction}
\label{Sec:Intro}

Theoretical and experimental investigation of Josephson junctions (JJs) made of unconventional superconductors showed that one can get so-called $\pi$ Josephson junctions\cite{Bulaevskii:pi-loop}, for which the first Josephson relation has the form $I_s=-I_c\sin\phi=I_c\sin(\phi+\pi)$, instead of $I_s=I_c\sin\phi$. There can be different reasons for this.
For example, in JJs formed using anisotropic superconductors with $d$-wave order parameter symmetry the additional phase shift of $\pi$ happens when charge carries enter the negative lobe of the order parameter and exit from the positive lobe. Experimental realizations of $\pi$-junctions include junctions formed at the boundary between two crystalline films of cuprate superconductors with different orientations\cite{Tsuei:Review}, ramp junctions between $d$-wave and $s$-wave superconductors\cite{Smilde:ZigzagPRL,Hilgenkamp:zigzag:SF} or JJs with ferromagnetic barrier \cite{Kontos:2002:SIFS-PiJJ,Ryazanov:2001:SFS-PiJJ}.
%
\iffalse In the superconductor-ferromagnet-superconductor (SFS) family of JJs the order parameter oscillates inside the F-layer, so that at S/F and F/S interfaces it may have different sign, resulting in a negative critical current\cite{Kontos:2002:SIFS-PiJJ,Ryazanov:2001:SFS-PiJJ}.\fi

If one considers a 1D long Josephson junction (LJJ) made of alternating 0 and $\pi$ parts, half-integer flux quanta (so-called semifluxons\cite{Goldobin:SF-Shape,Xu:SF-shape}) may spontaneously form at the joints between 0 and $\pi$ parts. Semifluxons were observed using SQUID microscopy in different types of 0-$\pi$-LJJs%
\cite{Hilgenkamp:zigzag:SF,Kirtley:SF:T-dep,%
Kirtley:SF:HTSGB}.

\iffalse
%SF.tech issues
Semifluxons are very interesting objects which have not been studied in detail, especially experimentally, first of all due to the relative difficulty to produce 0-$\pi$-LJJs. Only recently, the fabrication of high quality 0-$\pi$-LJJs based on \YBCO-Nb ramp zigzags has been reported\cite{Smilde:ZigzagPRL,Hilgenkamp:zigzag:SF}. Grain boundary junctions based on tri- and tetra-crystals are rather expensive and allow to study only a single semifluxon\cite{Kirtley:SF:T-dep}. Both 0 and $\pi$ junctions from the SFS family were demonstrated by several groups, however no SFS-LJJs with both 0 and $\pi$ sections have been realized so far to study semifluxons.
\fi

%SF.why interesting physics?
Semifluxons are very interesting objects which have not been studied in detail, especially experimentally. They are very different from integer fluxons. Fluxons are solitons, but semifluxons are not --- they are spontaneously formed and pinned at the 0-$\pi$-boundary. Consequently a semifluxon of either polarity represents the ground state of the system, while a fluxon moving in the LJJ always represents an excited state. This makes semifluxons attractive \eg{} for information storage and processing in classical and quantum regimes. In spite of pinning, semifluxons are able to hop from one 0-$\pi$-boundary to the other provided that they are not very far from each other. A chain of semifluxons usually exists in the antiferromagnetic ground state but can be ``polarized'' by means of an applied dc bias current\cite{Goldobin:SF-ReArrange}. A semifluxon has an eigen-mode oscillation frequency, so a 1D array of semifluxons has an energy band similar to the one in real crystals. By controlling the spacing between semifluxons (length of 0 and $\pi$-pieces) one can control the properties of such an artificial 1D crystal.

In this letter we propose and implement LJJs based on \emph{conventional} superconductors which allow one to study \emph{arbitrary} fractional vortices\footnote{Arbitrary fractional vortices may also appear in a LJJ with a strong second harmonic in the current-phase relation, which can be present either intrinsically or due to a faceted grain-boundary (frequently alternating regions of size $\ll\lambda_J$ of negative and positive critical current). In contrast to such ``splintered'' vortices which were also observed experimentally\cite{Mints:2002:SplinteredVortices@GB}, the fractional vortices discussed here are pinned at a single 0-$\pi$-boundary.} and, in particular, their \emph{dynamics}. Up to now this is not possible using natural 0-$\pi$-junctions which typically have a Stewart-McCumber parameter $\beta_c\sim5$ or even smaller. Having much lower damping, we succeeded to observe half-integer zero-field steps in the current-voltage characteristics which appears due to a non-trivial effect --- hopping of semifluxons between neighboring 0-$\pi$-boundaries. Low damping is also important in terms of future observations of macroscopic quantum effects involving fractional quanta\cite{Kato:1997:QuTunnel0pi0JJ}.

\section{The Model}
\label{Sec:Model}

The dynamics of the Josephson phase in LJJs consisting of alternating 0 and $\pi$ parts can be described by the 1D perturbed sine-Gordon equation\cite{Goldobin:SF-Shape}
\begin{equation}
  \phi_{xx}-\phi_{tt}-\sin\phi = \alpha\phi_t-\gamma-\theta_{xx}(x)
  , \label{Eq:sG-phi}
\end{equation}
where $\phi(x,t)$ is the Josephson phase and subscripts $x$ and $t$ denote the derivatives with respect to coordinate $x$ and time $t$. In Eq.~(\ref{Eq:sG-phi}) the spatial coordinate is normalized to the Josephson penetration depth $\lambda_J$ and the time is normalized to the inverse plasma frequency $\omega_p^{-1}$; $\alpha=1/\sqrt{\beta_c}$ is the dimensionless damping; $\gamma=j/j_c$ is the external bias current density normalized to the critical current density of the junction. The function $\theta(x)$ is a step function which is $\pi$-discontinuous at all points where 0 and $\pi$ parts join. 
%and is constant equal to $\pi n$ within each part. For example, 
%$\theta(x)$ can be equal to zero along all 0-parts and $\pi$ along 
%all $\pi$-parts.
According to Eq.~(\ref{Eq:sG-phi}), $\phi(x)$ is also $\pi$-discontinuous at the same points as $\theta(x)$. We call these points \emph{phase discontinuity points}.

To describe 0-$\pi$-LJJs one can use directly the equation with alternating critical current density%
\cite{Xu:SF-shape,Smilde:ZigzagPRL,Kirtley:IcH-PiLJJ,Buzdin:0-phi-LJJ} 
\begin{equation}
  \mu_{xx}-\mu_{tt}\pm\sin\mu = \alpha\mu_t-\gamma(x)
  \label{Eq:sG-mu}
\end{equation}
written for the \emph{continuous} phase $\mu(x,t)=\phi(x,t)-\theta(x)$.
Equations~(\ref{Eq:sG-phi}) and (\ref{Eq:sG-mu}) are equivalent\cite{Goldobin:SF-Shape}\footnote{For a single semifluxon the difference between $\phi$ and $\mu$ is demonstrated in Fig.~2 of Ref.~\onlinecite{Goldobin:SF-Shape}}.

\begin{figure}[tb]
  \centering\scalebox{0.8}{\includegraphics{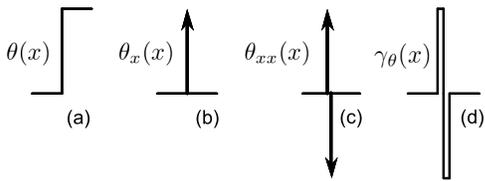}}
  \caption{
   $\theta(x)$, $\theta_{x}(x)$ and $\theta_{xx}(x)$ and its approximation $\gamma_{\theta}(x)$ by two rectangular pulses of the width $\Delta w$ and amplitude $\pi/\Delta w^2$.
  }
  \label{Fig:principle}
\end{figure}

Is it possible to create a 0-$\pi$-LJJ (LJJ with $\pi$-discontinuity) using a conventional LJJ? The fact that a junction is 0-$\pi$ is contained in the $\theta_{xx}(x)$ term. If $\theta(x)$ is a set of steps, then $\theta_{xx}(x)$ is a set of $-\delta(x)/x$ singularities as shown in Fig.~\ref{Fig:principle}(a)--(c). If we do not have initially the $\theta_{xx}$ term in Eq.~(\ref{Eq:sG-phi}), we can substitute its effect by introducing a properly shaped bias current $\gamma_{\theta}(x)=\theta_{xx}(x)$ in addition to the main bias current $\gamma$. To mimic one $\pi$-discontinuity the $\gamma_{\theta}(x)$ profile should be the one shown in Fig.~\ref{Fig:principle}(c). For the sake of practical treatment, one can approximate it by two rectangular pulses, as shown in Fig.~\ref{Fig:principle}(d). Such a bias profile can be created by two current injectors of width $\Delta w$ situated as close as possible to each other. The current of the amplitude $I_{\rm inj}=\pi I_c\lambda_J/\Delta w$ flowing from one injector to another should create a $\pi$-discontinuity of the Josephson phase. Thus one can create an artificial 0-$\pi$-LJJ provided that the width of the whole injector construction is well below $\lambda_J$. Note that, passing different currents, one can create \emph{arbitrary} $\kappa$-discontinuities instead of a $\pi$-discontinuity, and study arbitrary fractional vortices, if they appear\cite{Goldobin:SF-ReArrange}. This concept is a generalization of the idea to use a pair of injectors to insert a fluxon ($2\pi$-discontinuity) in an annular LJJ\cite{Ustinov:2002:ALJJ:InsFluxon}.

In conventional LJJs the phase $\phi=\mu$ is continuous while $I_{\rm inj}=0$. When we increase $I_{\rm inj}$ the phase $\phi$ develops a jump $\kappa\propto I_{\rm inj}$ at $x=x_{\rm inj}$. In practice, instead of a jump we have a rapid increase of the phase from $\phi$ to $\phi+\kappa$ over a small but finite distance. Physically this means that by passing a rather large current through the piece of the top electrode between two injectors we ``twist'' the phase $\phi$ by $\kappa$ over this small distance, \ie, we insert a magnetic flux equal to $\kappa\Phi_0/(2\pi)$ in a small distance between injectors. This flux has a characteristic size of the order of the distance between injectors or of their width. The junction may react on the appearance of such a discontinuity by forming a solution $\phi(x)$ [or $\mu(x)$] which changes on the characteristic length $\lambda_J$ outside the injector area, and corresponds, \eg, to the formation of a fractional Josephson vortex, as discussed below.

In the experiment the injectors are never ideal due to technological constraints and one has to calibrate them to know the value of current needed to get, \eg, a $\pi$-discontinuity. We do this by measuring the dependence of the critical current $I_c$ of the LJJ on the current through the injectors $I_{\rm inj}$, and comparing the obtained $I_c(I_{\rm inj})$ dependence with the $I_c(\kappa)$ dependence calculated theoretically. As a result we get $I_{\rm inj}/\kappa$.

\begin{figure}[tb]
  \centering\scalebox{0.8}{\includegraphics{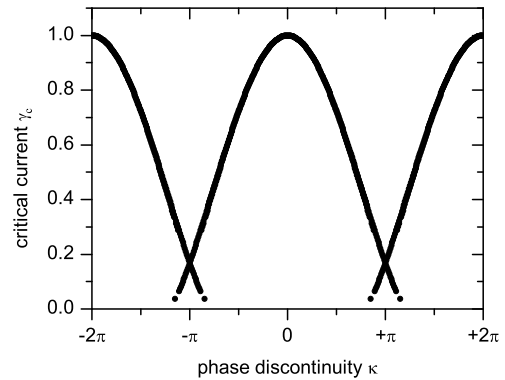}}
  \caption{
  Normalized critical current $\gamma_c$ \vs{} the Josephson phase discontinuity $\kappa\propto I_{\rm inj}$, numerically calculated for $L=2\lambda_J$.
  }
  \label{Fig:Ic(kappa)}
\end{figure}

The $\gamma_c(\kappa)$ curve ($\gamma_c$ is the maximum normalized supercurrent) numerically calculated for $L=2\lambda_J$ (like in our samples) is shown in Fig.~\ref{Fig:Ic(kappa)}. Details on the calculation of $\gamma_c(\kappa)$ for different $L$ will be given elsewhere\cite{Goldobin:LJJ:Ic(Iinj)}. It is important to note that $\gamma_c(\kappa)$ has maxima at $\kappa=2\pi n$ and cusp-like minima (possibly with hysteresis) at $\kappa=\pi(2n+1)$. The value of $\gamma_c$ at the minimum depends on the junction length and varies from 0 for a short junction to $2/\pi$ for an infinitely long one\cite{Goldobin:LJJ:Ic(Iinj)}.

\section{Experiment}
\label{Sec:ExpRes}

\subsection{Implementation}

\begin{figure}[b]
  \centering\includegraphics*{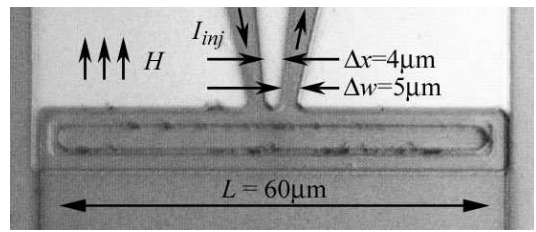}
  \caption{
    Optical microscope picture of the LJJ with two injectors.
    The width of the LJJ is $w=5\units{\mu m}$.
  }
  \label{Fig:Picture}
\end{figure}

The samples were fabricated\footnote{Hypres, Elmsford (NY), USA. http://www.hypres.com} using standard Nb-AlO$_x$-Nb technology with low critical current density $j_c\approx 100\units{A/cm^2}$ to have  $\lambda_J\sim30\units{\mu m}$ as large as possible in comparison with the injector size. Data reported here were obtained from the sample shown in Fig.~\ref{Fig:Picture} at $T\approx5\units{K}$.%KL609-156

The measured $I$-$V$ characteristics (IVCs) and $I_c(H)$ dependences (presented below) ensure good sample quality and the absence of trapped magnetic flux.

\begin{figure}[tb]
  \centering\scalebox{0.8}{\includegraphics{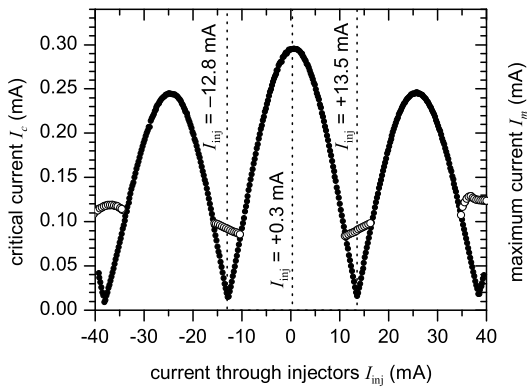}}
  \caption{
    Experimentally measured dependence of $I_c(I_{\rm inj})$ (full circles) and the maximum current of the semi-integer zero field steps $I_m(I_{\rm inj})$ (open circles) at $H=0$. The dependences are perfectly mirror symmetric relative to the $I_{\rm inj}$ axis.
  }
  \label{Fig:Ic(Iinj)}
\end{figure}

To calibrate the injectors, we measured the dependence $I_c(I_{\rm inj})$ at $H=0$ shown in Fig.~\ref{Fig:Ic(Iinj)}. Obviously, the dependence is similar to Fig.~\ref{Fig:Ic(kappa)}, and the minimum $I_c$ is about $10\units{\%}$ of the maximum $I_{c}$, in agreement with numerical calculations for $L/\lambda_J=2$. From Fig.~\ref{Fig:Ic(Iinj)} we find that $\kappa=+\pi$ corresponds to $I_{\rm inj}\approx+13.5\units{mA}$ and $\kappa=-\pi$ corresponds to $I_{\rm inj}\approx-12.8\units{mA}$. The pattern is slightly shifted along the $I_{\rm inj}$ axis probably due to self-field effects. Thus, the maximum $I_c$ is reached at $I_{\rm inj}\approx+0.3\units{mA}$.

The value of $I_c$ at the maxima slowly decreases with increasing $|I_{\rm inj}|$. This effect can be reproduced in numerical simulations using the model (\ref{Eq:sG-phi}) for injectors with finite $\Delta x$ and $\Delta w$ like in our samples.

\begin{figure}[b]
  \centering\scalebox{0.8}{\includegraphics{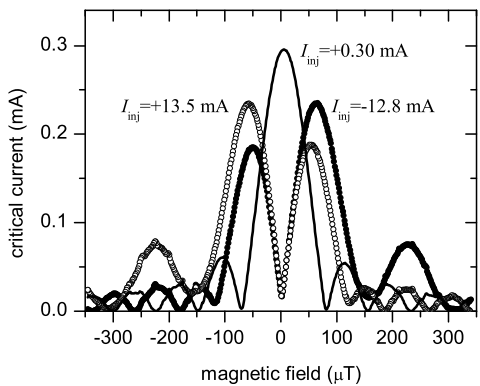}}
  \caption{
    Experimentally measured dependence $+I_c(H)$ for 0$|$0 and $+\pi|$0 and $-\pi|$0 states of the junction. The dependence of the negative critical current $-I_c(H)$ is not shown, but it is perfectly mirror symmetric with respect to the $H$-axis.
  }
  \label{Fig:Ic(H)}
\end{figure}

One of the specific features of natural 0-$\pi$-junctions is the minimum in the $I_c(H)$ dependence at $H=0$\cite{Smilde:ZigzagPRL}. It was calculated\cite{Kirtley:IcH-PiLJJ} that when the 0-$\pi$-junction is short ($L\ll\lambda_J$) the minimum is very deep and $I_c$ reaches zero at $H=0$. As $L$ increases, the minimum gets more shallow and at $L\sim10\lambda_J$ almost disappears. In our artificial 0-$\pi$-LJJ with $L\approx2\lambda_J$ the minimum should be strongly developed. We have measured and compared $I_c(H)$ at $\kappa=0$ (no discontinuity) and at $I_c(H)$ at $\kappa=\pm\pi$ ($I_{\rm inj}=+13.5\units{mA}$ and $I_{\rm inj}=-12.8\units{mA}$). The results are presented in Fig.~\ref{Fig:Ic(H)} and are in good agreement with the prediction\cite{Kirtley:IcH-PiLJJ}. One can notice that at each value of $I_{\rm inj}$ corresponding to a 0-$\pi$-LJJ, the left and right lobes of the $I_c(H)$ curve look somewhat asymmetric with the period of oscillations being different for $H>0$ and $H<0$. Again, this is related to the finite distance and width of the injectors and can be reproduced in simulations. A similar asymmetry has been calculated in Ref.~\onlinecite{VanHarlingen:1995:Review} (Fig.~12, the last $s$-wave plot) where the ``inserted'' flux is due to a trapped vortex, coupling $\Phi_0/2$ to the center of a conventional junction. If the size of the trapped vortex (in our case $\Delta w$ and $\Delta x$) approaches zero, $I_c(H)$ approaches a perfectly symmetric shape with equal oscillation periods at $H>0$ and $H<0$ (see Fig.~12 of Ref.~\onlinecite{VanHarlingen:1995:Review}, the first $d$-wave plot).

\iffalse
We would like to mention that it is also possible to create a 0-$\pi$-junction just using, instead of the injectors, a single Abrikosov vortex which couples half a flux quantum to the middle of the junction and has the size $\ll\lambda_J$. It was argued\cite{VanHarlingen:1995:Review} that one can distinguish between the case of a ``real'' 0-$\pi$-junction and the case of a conventional JJ with trapped Abrikosov vortex (see Fig.~12 of Ref.~\onlinecite{VanHarlingen:1995:Review}, {\cf} the last plot for $s$-wave and the first plot for $d$-wave). We argue that such a distinction can be done only if the size of the vortex is not negligible in comparison with $\lambda_J$ or $L$. When the size of vortex/injectors is $\ll(\lambda_J,L)$, no distinction can be made. In fact, in this limit both systems are formally equivalent.
\fi

\subsection{Semi-integer zero-field steps}

\begin{figure}[tb]
  \centering\scalebox{0.8}{\includegraphics{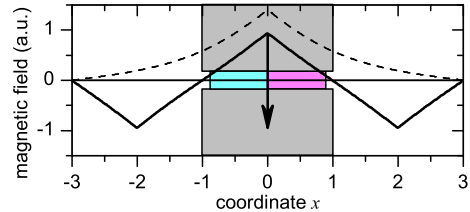}}
  \caption{
    Sketch of a semifluxon in a 0-$\pi$-LJJ and two images outside the junction. Dashed line shows how the magnetic field of semifluxon would look like in an infinite LJJ. Solid line shows the magnetic field of a semifluxon in a LJJ of $L/\lambda_J=2$. The arrow indicates the localized field produced by the injector.
  }
  \label{Fig:sketch}
\end{figure}

When a LJJ has a $\pi$-discontinuity in its center, the ground state consists of a semifluxon spontaneously formed and pinned at the $\pi$-discontinuity\cite{Kirtley:IcH-PiLJJ,Goldobin:SF-Shape,Goldobin:SF-ReArrange}, as shown in Fig.~\ref{Fig:sketch}. When $L\sim\lambda_J$ the semifluxon solution\cite{Xu:SF-shape,Goldobin:SF-Shape}, shown in Fig.~\ref{Fig:sketch} by a dashed line, is not valid anymore as it does not satisfy the boundary conditions $\phi_x(\pm L/2)=0$. The semifluxon solution derived in Refs.~\onlinecite{Xu:SF-shape,Goldobin:SF-Shape} is valid in an infinite LJJ only. To construct the solution for a LJJ of finite length, we introduce images (anti-semifluxons) situated outside the left and the right edge of the junction at a distance of $L/2$ from each edge\footnote{To balance the problem completely, one needs to introduce an infinite chain of images: positive semifluxons at $x=2nL$ and negative semifluxons at $x=(2n+1)L$.}. The final magnetic field profile $\phi_x(x)$ is shown in Fig.~\ref{Fig:sketch} by the thick solid line.

If one applies a bias current to a chain of antiferromagnetically ordered semifluxons, the semifluxons can rearrange\cite{Goldobin:SF-ReArrange}: the positive semifluxon at $x=0$ hops to the $\pi$-discontinuity at $x=L$ and the image hops from $x=L$ to $x=0$. This process is accompanied by a transfer of flux equal to $\Phi_0$ across the right edge of the junction. Subsequently the same mutual hopping happens at the left edge of the junction and the flux equal to $\Phi_0$ is transferred across the left edge too. This is repeated periodically in time, and the net transferred flux results in a finite voltage across the junction. This process is very similar to the dynamics which takes place in a LJJ biased at the zero-field step (ZFS)%
\footnote{
  Since our LJJ has intermediate length, we speak here about ZFS which are better described by a multi-mode approach\cite{Enpuku:MultiModeZFS} rather than by perturbation theory\cite{Pedersen:ZFS-cmp,Levring:ZFS-ov-in,Levring:ZFS(H)-ov-in}
}.

\begin{figure}[tb]
  \centering\scalebox{0.8}{\includegraphics{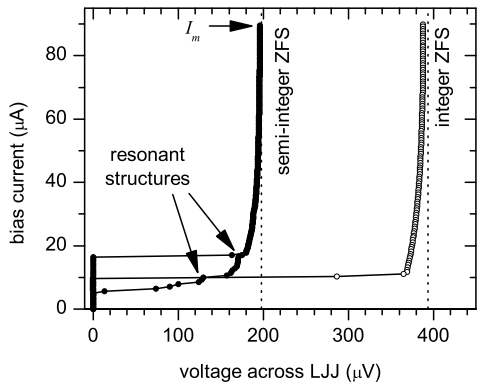}}
  \caption{
    Experimentally measured integer and semi-integer zero field steps on the IVC of the artificial 0-$\pi$-LJJ at $I_{\rm inj}=+13.5\units{mA}$ (full circles) and $I_{\rm inj}=0$ (open circles).
  }
  \label{Fig:ZFS}
\end{figure}

Presuming that the flux in our 0-$\pi$-LJJ is transferred with the same maximum velocity, but only $\Phi_0$ rather $2\Phi_0$ per hopping/reflection, we should observe half-integer ZFS on the IVC of the 0-$\pi$-LJJ, as was recently predicted by Stefanakis\cite{Stefanakis:ZFS/2}. Note that to observe semi-integer ZFS the 0-$\pi$-LJJ should be not very long ($L\lesssim 5\lambda_J$) so that the semifluxons can still hop over this distance.

In Fig.~\ref{Fig:ZFS} we show two IVCs taken at $H=0$: one with a classical ZFS recorded at $I_{\rm inj}=0$ and another with a semi-integer ZFS registered at $I_{\rm inj}=+13.5\units{mA}$. The asymptotic voltage of the ZFS is $V_{\rm ZFS}=388\units{\mu V}$, while for the semi-integer ZFS it is $V_{\rm ZFS/2}=196\units{\mu V}$. Thus, $2V_{\rm ZFS/2}\approx V_{\rm ZFS}$ with accuracy of $1\units{\%}$. This is the experimental confirmation of the existence of semi-integer ZFS\cite{Stefanakis:ZFS/2} and semifluxon hopping\cite{Goldobin:SF-ReArrange}. We also observe resonant structures at the half-integer ZFS, which is similar to structures we see in simulations. This kind of resonances typically appear when magnetic flux moves in a LJJ with inhomogeneities\cite{Ustinov:inhomo}.

We also measured the behavior of the semi-integer ZFS as a function of $I_{\rm inj}$ and found that \emph{the asymptotic voltage of the semi-integer ZFS does not depend on $I_{\rm inj}$}. When $\kappa\ne\pi$, one has two types of fractional vortices in the system: a vortex $\kappa$ at $x=0$ and two images (antivortices) $\kappa-2\pi$ at $x=\pm L$. When the vortex and antivortex hop and exchange positions, the total flux transferred is still $\Phi_0$ (they exchange a virtual fluxon\cite{Goldobin:SF-ReArrange}). Although the voltage of the semi-integer ZFS is not affected by $I_{\rm inj}$, the amplitude of the step $I_{\max}$ changes with $I_{\rm inj}$ until the step gets completely hidden by the critical current. The dependence of the maximum current of the semi-integer ZFS on $I_{\rm inj}$ is shown in Fig.~\ref{Fig:Ic(Iinj)}. No high order ZFSs were observed since our LJJ is rather short ($\sim2\lambda_J$).

\section{Conclusions}
\label{Sec:Conclusion}

In conclusion, we have proposed, implemented and tested experimentally long Josephson junctions based on Nb-AlO$_x$-Nb technology with an arbitrary $\kappa$-discontinuity of the Josephson phase created by passing a current through a pair of injectors. Formally this system is equivalent to a 0-$\pi$ long Josephson junction made of unconventional superconductors and allows one to study the \emph{dynamics} of fractional vortices. We confirmed the hopping of semifluxons under the action of applied bias current which manifests itself as the appearance of semi-integer zero field steps in the current-voltage characteristics of the junctions.

\iffalse
  This model system helps to shed light on the physics of fractional vortices in LJJs made of unconventional superconductors. It also enables the investigation of naturally ``forbidden'' configurations such as one $\pi$-discontinuity in an annular long Josephson junction.
\fi

Among the many interesting physical questions concerning the physics of fractional vortices we just mention two examples for which \emph{arbitrary quantization} and/or \emph{low damping} are essential: (1) How does the eigen-frequency of a fractional vortex depend on the flux carried by it? How does the coupling between neighboring fractional vortices affect their eigen-modes (eigen-mode splitting) and can one construct an artificial 1D crystal with controllable energy bands? (2) Does a semifluxon behave as a spin-$\frac12$ particle? Can it make quantum transitions $\uparrow\leftrightarrow\downarrow$ or be in a superposition of both states or does flux conservation prevent this? Can one create superpositions of states such as $\uparrow\downarrow\leftrightarrow\downarrow\uparrow$ for two coupled semifluxons? Finally, we note that an annular system with \eg{} three semifluxons may serve as an interesting system to study frustration effects.

\begin{acknowledgments}
  We acknowledge discussions with B. Chesca and support by the Deutsche Forschungsgemeinschaft, and by the ESF programs "Vortex" and "Pi-shift".
\end{acknowledgments}

\bibliography{LJJ,SF,QuComp,SFS,pi}

\end{document}